\documentclass[twocolumn,showpacs,prb]{revtex4} 
\usepackage{amssymb} 
\usepackage{graphicx}
\usepackage{subfigure}
\usepackage{dcolumn}
\usepackage{bm}
\begin{document}

\title{Phonon dispersions and vibrational properties of monolayer, bilayer, and trilayer graphene}
\author{Jia-An Yan, W. Y. Ruan, and M. Y. Chou}
\affiliation{School of Physics, Georgia Institute of Technology,
Atlanta, Georgia 30332-0430, U.S.A.}
\date{\today}
\begin{abstract}

The phonon dispersions of monolayer and few-layer graphene (AB
bilayer, ABA and ABC trilayers) are investigated using the
density-functional perturbation theory (DFPT). Compared with the
monolayer, the optical phonon $E_{2g}$ mode at $\Gamma$ splits into
two and three doubly degenerate branches for bilayer and trilayer
graphene, respectively, due to the weak interlayer coupling. These
modes are of various symmetry and exhibit different sensitivity to
either Raman or infrared (IR) measurements (or both). The splitting
is found to be 5 cm$^{-1}$ for bilayer and 2 to 5 cm$^{-1}$ for
trilayer graphene. The interlayer coupling is estimated to be about
2 cm$^{-1}$. We found that the highest optical modes at K move up by
about 12 cm$^{-1}$ for bilayer and 18 cm$^{-1}$ for trilayer
relative to monolayer graphene. The atomic displacements of these
optical eigenmodes are analyzed.
\end{abstract}


\maketitle

\section{Introduction}

In recent years, monolayer and few-layer graphene have attracted
great attention due to the unique properties observed
experimentally. \cite{DeHeer2007,Geim2007,Katsnelson2007} Many
intriguing transport phenomena, such as ballistic transport at room
temperature,\cite{Berger2004,Berger2006} the anomalous quantum Hall
effect, \cite{Novoselov2005, Zhang2005} and novel many-body
couplings \cite{Bostwick2006} have been reported. In addition to
being a physical system exhibiting novel properties, graphene and
graphene layers have been proposed as promising candidates for
future nanoelectronics. The epitaxial graphene grown on SiC is of
particular interest due to the compatibility with current silicon
technology. \cite{DeHeer2007,Hass2007}

Besides their unusual electronic structure, \cite{Latil2006}
vibrational properties and phonon spectra are also of fundamental
interest from which many physical properties (such as thermal
conductivity and heat capacity) can be derived. Furthermore, phonons
are crucial for studying the quasiparticle dynamics
\cite{Bostwick2006} and electrical transport properties. Electrons
excited by optical methods can be scattered into another states by
electron-phonon coupling (EPC). It has been suggested that the
scattering between electrons and the optical phonon modes greatly
affects the high-field ballistic transport properties in carbon
nanotubes. \cite{Lazzeri2005} In graphene and metallic single-wall
nanotubes, the EPC strongly affects the phonon frequencies, giving
rise to Kohn anomalies \cite{Kohn1959,Piscanec2004} and possible
soft modes or Peierls distortions. \cite{Dubay2003,Piscanec2007}

Many experimental methods have been used to measure the phonon
dispersions of graphite, such as inelastic neutron scattering (INS),
\cite{Nicklow1972} electron-energy loss spectroscopy
(EELS),\cite{Oshima1988} high-resolution electron energy-loss
spectroscopy (HREELS),\cite{Siebentritt1997} and inelastic x-ray
scattering (IXS). \cite{Maultzsch2004,Mohr2007} These measurements
require large enough samples of crystalline quality and are limited
to specific directions or phonon modes. More recently, Mohr et al.
\cite{Mohr2007} have presented complete measurements of both the
optical and the acoustic phonon modes along the directions
$\Gamma$-K-M-$\Gamma$ of graphite using IXS. The results in these
measurements are very close. In contrast to bulk graphite, Raman
scattering has been widely used for probing the G-band in graphene
layers that corresponds to the $\Gamma$ phonons.
\cite{Ferrari2006,Graf2006,Gupta2006,Yan2006} Recently reported
Raman spectra for graphene layers show that the intensity and
position of the first-order G-band as well as the second order
D-band (historically named the G' band) are modified with an
increasing number of layers. \cite{Ferrari2006, Graf2006}

On the theoretical side, Gr\"{u}neis et al. \cite{Gruneis2002}
presented the phonon dispersions of graphite using the
4th-nearest-neighbor force constant (4NNFC) approach. However, it
has been argued that due to the Kohn anomaly at $\Gamma$ and K, it
is not possible to obtain the correct phonon dispersions near
$\Gamma$ and K from the force constant method. \cite{Piscanec2004}
Dubay and Kresse \cite{Dubay2003} performed density-functional theory (DFT)
calculations of the phonon dispersions in graphite within the local
density approximation (LDA). Their results are in good agreement
with phonon-measurements by HREELS. Using the LDA and the generalized-gradient approximation
(GGA), Wirtz and Rubio \cite{Wirtz2004} calculated the phonon
dispersions of graphite and obtained results close to the vast
majority of the experimental data-points. At the GGA-PBE level,
Mounet and Marzari \cite{Mounet2005} also presented a detailed
calculation of the phonon dispersions of graphene and graphite.

With regard to graphene layers, it is unclear how the phonon
properties are affected by the stacking order and the weak
interlayer coupling. This effect is important for understanding the
EPC in multilayer graphene as well as the interpretation of the
Raman spectra. For example, the phonon dispersion around K is
crucial for the correct interpretation of the Raman second order D
peak. It has been shown that in few-layer graphene, the electronic
dispersions near the Fermi level exhibit various features depending
on the stacking order. \cite{Latil2006} In this work, the
vibrational properties of one- and few-layer graphenes are
calculated using density-functional perturbation theory (DFPT).
\cite{Baroni2001} The monolayer, bilayer (AB stacking), and trilayer
(ABA and ABC stackings) are considered in order to illustrate the
effects of stacking order and interlayer coupling. The van der Waals
corrections in graphite have been shown to be important in order to
correctly describe the long-range binding properties.
\cite{Girifalco2002, Kack2006, Dappe2006} However, previous
theoretical calculations based on DFT with both LDA \cite{Dubay2003}
and GGA \cite{ Mounet2005, Mohr2007} have indicated that rather
reasonable vibrational properties of graphite can be
obtained within DFT as compared with experiments. We find that the phonon dispersions for graphene
and graphene layers exhibit somewhat different characteristics,
especially at $\Gamma$ and K. Detailed analysis of the phonon modes
is also presented.

\section{Computational Details}

Density-functional calculations are performed using the ESPRESSO
code \cite{pwscf} with the LDA. Troullier-Martin (TM)
norm-conserving pseudopotentials \cite{Troullier1991} generated from
the valence configuration of $2s^22p^2$ for C are employed. The
wavefunction and the charge density are expanded using energy
cutoffs of 110 and 440 Ryd, respectively. Methfessel-Paxton smearing
\cite{Methfessel1989} with an energy width of 0.03 Ryd is adopted
for the self-consistent calculations. The dynamical matrices are
calculated based on DFPT within the linear response. For the
integration over electronic states in the calculations, we use a $48
\times 48\times 1$ uniform k-point mesh. A $6\times 6 \times 1$ grid
is used for the phonon calculation to obtain the dynamical matrices.
We have carefully tested these parameters and the phonon frequencies
are converged to be within 1 cm$^{-1}$.


\begin{figure}[tbp]
\centering
   \includegraphics[width=9cm]{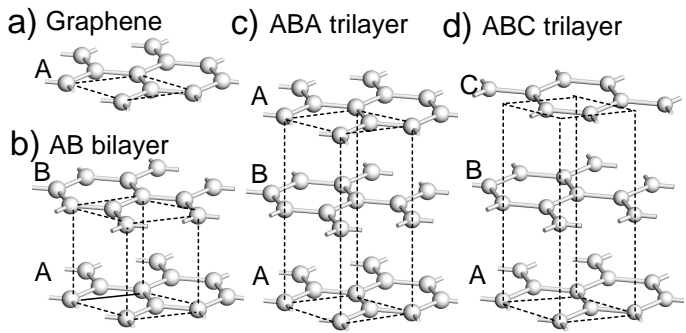}
  \caption{Stacking structure for (a) monolayer, (b) AB bilayer,
  (c) ABA trilayer and (d) ABC trilayer graphene. }
\label{fig:model}
\end{figure}

Figure~\ref{fig:model} shows the two-dimensional (2D) primitive cells
for monolayer, bilayer (AB stacking), and trilayer graphene (ABA and
ABC stacking). A large vacuum region of more than 10 \AA~ along the
$z$ direction is used to minimize the interactions between graphene
layers in different supercells.

\section{Results and Discussions}

\begin{figure}[tbp]
\centering
 \includegraphics[height=6.0cm,angle=270]{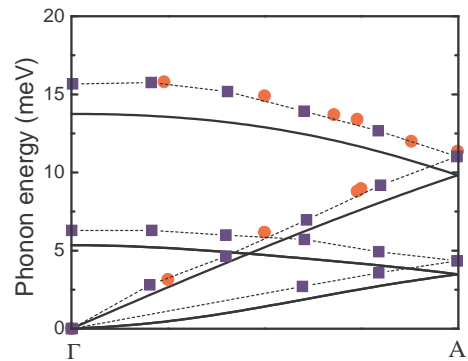}
\caption{(Color online) Phonon dispersions of graphite along the
$\Gamma$-A direction. Solid lines are present calculational results.
Circles are IXS data from Ref.~\onlinecite{Mohr2007}, and squares
are neutron scattering data from Ref.~\onlinecite{Nicklow1972}. The
dotted lines are smooth curves through the measured points. }
\label{fig:GA}
\end{figure}

The optimized LDA lattice constant in the graphene plane is 2.45
\AA, in good agreement with the previous calculated result,
\cite{Latil2006} which is also close to the experimental value of
2.46 \AA~for graphite. \cite{Wyckoff1963} For the bilayer and
trilayer systems, the lattice constant in the plane remains almost
the same as in graphene. The optimized interlayer spacing is 3.33
\AA, slightly smaller than the experimental value of 3.35 \AA~in
bulk graphite. \cite{Wyckoff1963} In comparison, we obtained a
theoretical value of 3.32 \AA~for graphite, which is close to
previous LDA results. \cite{Ooi2006} The interlayer binding energy
(defined as the total energy difference between the coupled and
uncoupled graphene layers) of bulk graphite is calculated to be 25.2
meV/atom, while this energy falls to 12.3 meV/atom for an AB
bilayer. For the ABA and ABC trilayer, the interlayer binding
energies are both 16.5 meV/atom. Our result of graphite is
comparable to previous calculations using a combined
density-functional and intermolecular perturbation theory approach.
\cite{Dappe2006} To validate the DFPT phonon calculations, we also
calculated the low energy phonon dispersions in bulk graphite along
the $\Gamma$-A direction (perpendicular to the layers), as shown in
Fig.~\ref{fig:GA}. Except for a small frequency shift, our LDA
dispersions agree with experimental data rather well. Therefore, we
believe our calculations yield reliable descriptions of the phonon
properties in graphite and graphene layers. This agreement between
LDA and experimental results indicate an error cancellation for
energy variations near the equilibrium layer separation, even though
the state-of-the-art local or semilocal exchange-correlation
functionals are not able to properly describe the long-range
interlayer interactions dominated by van der Waals dispersion forces
(see Refs. [\onlinecite{Kohn1998}] and [\onlinecite{Rydberg2003}]
for details).


\subsection{Phonon Properties of Graphene\label{sec:A}}

\begin{figure}[tbp]
\centering
 \includegraphics[width=7.5cm]{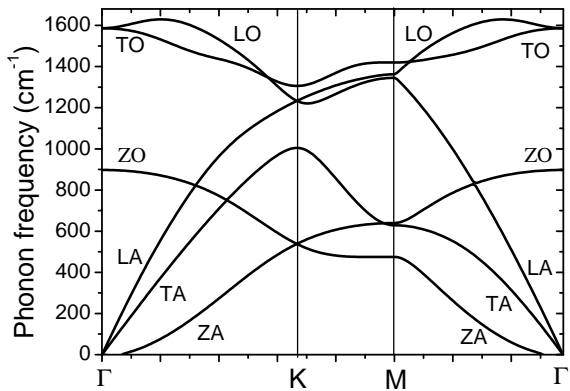}
\caption{Phonon dispersions for monolayer graphene. } \label{fig:mono-ph}
\end{figure}

Figure~\ref{fig:mono-ph} shows the phonon dispersions for monolayer
graphene calculated at the theoretical lattice constant, which will
be compared with multilayer results in the next section. In contrast
to the linear dispersion near the $\Gamma$ point for the in-plane TA
and LA modes, the out-of-plane ZA mode shows a $q^2$ dispersion,
which is a characteristic feature of the phonon dispersions in
layered crystals as observed experimentally. \cite{Mounet2005,
Zabel2001, Lifshitz1952} The same feature also appears in bilayer
and trilayer phonon dispersions, as will be discussed in Section
\ref{sec:B}.

The calculated frequency (1586 cm$^{-1}$) of the degenerate LO and
TO modes at $\Gamma$ is slightly smaller than the previous value of
1595 cm$^{-1}$ obtained by Dubay \emph{et al.},\cite{Dubay2003} but
is in excellent agreement with the experimental result of 1587
cm$^{-1}$ by inelastic x-ray scattering measurements.
\cite{Maultzsch2004} At the Brillouin zone corner K, the phonon
energy of the symmetric TO $A_1'$ mode (1306 cm$^{-1}$) is close to
the frequency (1326 cm$^{-1}$) calculated by Wirtz et al.
\cite{Wirtz2004} Our result is also consistent with the estimate by
Yao et al. \cite{Yao2000} from their high-voltage transport
measurements for graphite. They suggested that scattering by phonons
with an energy of about 1300 cm$^{-1}$ gives rise to the dramatic
conductance drop at a high bias. In general, our calculated phonon
dispersions for monolayer graphene are comparable with those
obtained in previous calculations \cite{Dubay2003,Wirtz2004} and
agree very well with experimental results.
\cite{Maultzsch2004,Yao2000}


A previous study by Piscanec \emph{et al.} \cite{Piscanec2004}
showed that the degenerate $E_{2g}$ modes at $\Gamma$ and the
highest TO mode at K have a strong EPC, leading to Kohn anomalies in
the phonon dispersions. A detailed analysis of the origin of the
strong EPC for these modes is presented below.

For a specific phonon mode $\nu$ with wave vector $q$, the
displacement of atom $j$ ($j$=$\alpha$,$\beta$) in unit cell $m$
will oscillate according to the following expression in the
classical picture:
\begin{eqnarray}
\nonumber \vec{u}_{q\nu}^{mj}&=&\sum_{s=x,y,z} \hat{e}_s
\mathrm{Re}\{\epsilon_{q\nu,s}^j
e^{i(\vec{q}\cdot\vec{R}_m-\omega_{q\nu} t)}\}\\
& = &\sum_{s=x,y,z} \hat{e}_s |\epsilon_{q\nu,s}^j|
\mathrm{cos}(\vec{q}\cdot\vec{R}_m-\omega_{q\nu} t+\phi_{q\nu,s}^j),
\end{eqnarray}
where $\vec{R}_m$ is the lattice vector for unit cell $m$,
$\phi_{q\nu,s}^{j}$ denotes the phase factor of the complex
eigenvector $\epsilon_{q\nu,s}^j$, and
$\omega_{q\nu}$ is the phonon frequency.

\begin{figure}[tbp]
\centering
 \includegraphics[width=8.0cm]{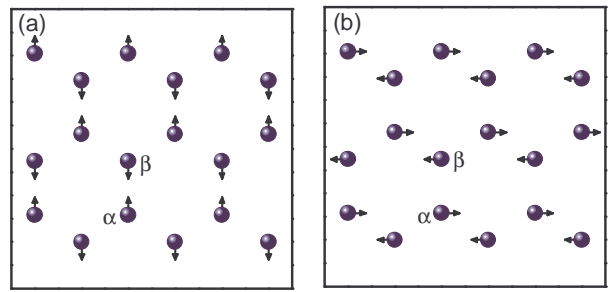}
\caption{(Color online) Pattern of atomic displacements for the
TO/LO modes at $\Gamma$ in monolayer graphene. } \label{fig:E2g}
\end{figure}

For the degenerate TO/LO phonon modes at $\Gamma$,
Figs.~\ref{fig:E2g}(a) and (b) schematically show the atomic
displacements associated with the two eigenmodes. Clearly, two
neighboring atoms vibrate opposite to one another. This gives rise
to a large bond distortion and couples to electronic states near the
Dirac point (which can be projected into two states localized at atom $\alpha$ and $\beta$, respectively) through an intravalley scattering (with phonon
$q$$\approx$0). Therefore, a strong EPC is expected, which has also
been demonstrated by the effective mass theory \cite{Ando2006} as
well as the tight-binding model. \cite{Jiang2004}


\begin{figure}[tbp]
\centering
 \includegraphics[width=8.5cm]{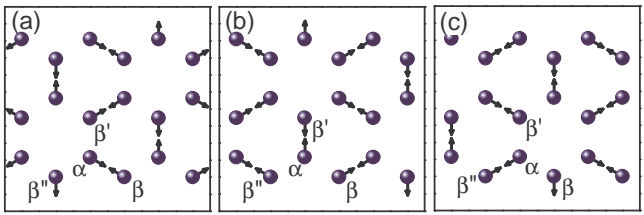}
\caption{(Color online) Three snapshots of atomic displacements for
the highest TO mode at K in monolayer graphene, where atom $\alpha$
approaches its three nearest neighbors $\beta$ (a), $\beta'$ (b),
and $\beta''$ (c) successively. } \label{fig:K6}
\end{figure}

In contrast, for the highest TO A$_1'$ mode at K the classical
displacements of neighboring atoms $\alpha$ and $\beta$ follow the
pattern
\begin{eqnarray}
\vec{u}^{\alpha} & = &u_0[\hat{e}_x
\mathrm{cos}(\frac{\pi}{2}-\omega t)-\hat{e}_y
\mathrm{sin}(\frac{\pi}{2}-\omega t)],\\
\vec{u}^{\beta} & = &u_0[\hat{e}_x \mathrm{cos}(\frac{\pi}{2}-\omega
t)+\hat{e}_y \mathrm{sin}(\frac{\pi}{2}-\omega t)]
\end{eqnarray}
and atoms $\alpha$ and $\beta$ move circularly. In particular, one
moves counterclockwise, while the other clockwise, as shown in
Fig.~\ref{fig:K6}. Accordingly, each atom approaches its three
nearest neighbors successively during one period. Fig.~\ref{fig:K6}
shows three snapshots of the atomic displacements in one period.


Since the degenerate electronic states at the Dirac point can be
projected into two states localized at atom $\alpha$ and $\beta$,
respectively, the above mode of ionic vibration facilitates the
transition of an electron from atoms $\alpha$ to $\beta$ or vice
versa, resulting in the electronic intervalley scattering via a
phonon with momentum $\vec{K}$. Therefore, a strong EPC is expected
for this mode compared with other modes at K. \cite{Lazzeri2006prl}
Based on the detailed analyses of these modes, we anticipate a
distinct electron-phonon interactions for these modes in few-layer
graphene. The results will be presented elsewhere.

\subsection{Phonon Dispersions for Graphene Layers\label{sec:B}}

\begin{figure*}[tbp]
\centering
 \includegraphics{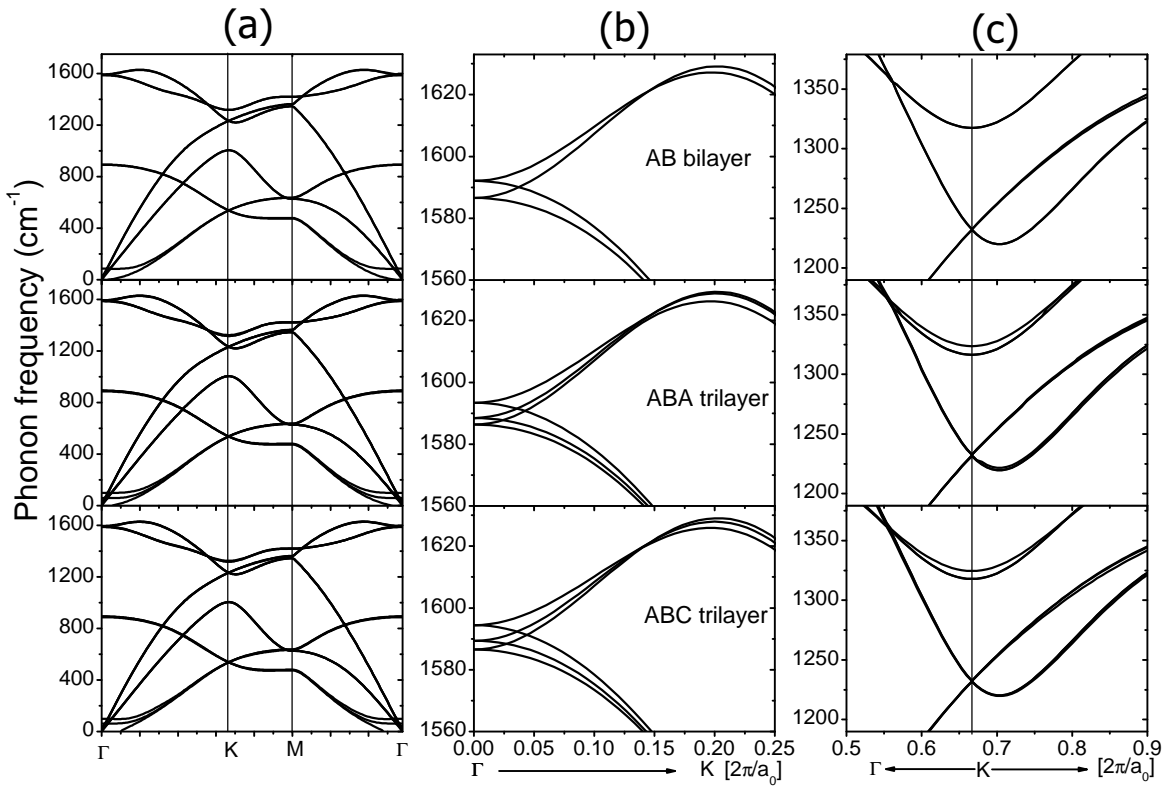}
\caption{Phonon dispersions for graphene multilayers. From top to
bottom: AB bilayer, ABA trilayer, and ABC trilayer. Column (a): full
phonon spectra; column (b): optical phonon dispersions near
$\Gamma$; column (c): optical phonon dispersions near K.}
\label{fig:layer-ph}
\end{figure*}

In this section, we focus on the optical phonon modes in multilayer
graphene. The phonon dispersions for bilayer and trilayer are shown
in Fig.~\ref{fig:layer-ph} (a). The detailed dispersions for the
high optical branches near $\Gamma$ and K are enlarged in
Figs.~\ref{fig:layer-ph}(b) and (c), respectively. The optical
phonon frequencies are also listed in Table~\ref{tab:freq}.

\begin{table*}[tbp]
 \caption{High optical phonon frequencies $\omega$ (in cm$^{-1}$) at $\Gamma$ and K for
 monolayer, bilayer, trilayer graphene, and bulk graphite. The phonon frequencies at $\Gamma$ and K from
recent DFT calculations (with LDA and GGA) as well as experimental
measurements are listed for comparison. The point group symmetry at
$\Gamma$ (K) for the monolayer, AB bilayer, ABA trilayer, ABC
trilayer, and graphite is $D_{6h}$ ($D_{3h}$), $D_{3d}$ ($C_{3v}$),
$D_{3h}$ ($C_{3h}$), $D_{3d}$ ($C_{3v}$), and $D_{6h}$ ($D_{3h}$),
respectively. In parentheses are the mode symmetries. }
\label{tab:freq}
\begin{ruledtabular}
\begin{tabular}{cllllll}

            & Graphene   &   AB      &   ABA   & ABC & Graphite & Graphite exp.\\
\hline
$\Gamma$    & 1586 ($E_{2g}$)  &   1587 ($E_{g}$)   &   1586 ($E'$) &  1586 ($E_g$) & 1586 ($E_{2g}$)&  1582\footnotemark[6],1581\footnotemark[8]\\
            & 1595\footnotemark[1], 1597\footnotemark[2]\footnotemark[3]                & 1592 ($E_u$)   &   1588 ($E''$)  &  1589 ($E_{u}$)  & 1595 ($E_{1u}$) & 1588\footnotemark[7]   \\
            & 1569\footnotemark[4], 1581\footnotemark[5]     &     &  1593 ($E'$)     & 1594 ($E_g$)    &  &     \\
            &            &           &        &            &      & \\
K           & 1306 ($A_1'$)       &   1318 ($E$)     &   1316 ($E_1'$, $E_1''$)    &  1318 ($E$) &  1322 ($E$)  & \\
            & 1371\footnotemark[1], 1326\footnotemark[3]&  & 1324 ($E_2'$)    &  1325 ($A_1$)    &      & \\
            & 1289\footnotemark[4], 1300\footnotemark[5]&            &         &           &       &\\
            & 1265\footnotemark[8]          &           &         &       &  &\\
\end{tabular}
\end{ruledtabular}
\footnotetext[1]{LDA, soft projector augmented wave (PAW), Ref.
[\onlinecite{Dubay2003}].} \footnotetext[2]{LDA, Hard PAW, Ref.
[\onlinecite{Dubay2003}].} \footnotetext[3]{LDA, TM potentials, Ref.
[\onlinecite{Wirtz2004}].} \footnotetext[4]{GGA, TM potentials, Ref.
[\onlinecite{Wirtz2004}].} \footnotetext[5]{GGA, Ref.
[\onlinecite{Maultzsch2004}].} \footnotetext[6]{ Expt.
$\omega(E_{2g})$,
Refs.[\onlinecite{Nemanich1979,Touninstra1970,Brillson1997}].}
\footnotetext[7]{ Expt. $\omega(E_{1u})$, Ref.
[\onlinecite{Nemanich1979}] and [\onlinecite{Friedel1971}].}
\footnotetext[8]{Inelastic X-ray data of Ref.
[\onlinecite{Maultzsch2004}] and [\onlinecite{Mohr2007}].}
\end{table*}

Compared with the monolayer result, several distinct features can be
identified for graphene multilayers. First, there is one (two)
additional low-frequency mode with energy of about 90 cm$^{-1}$ at
$\Gamma$ in bilayer (trilayer) graphene. These modes arise from
interlayer movement (so-called `layer breathing' modes). Second, at
$\Gamma$ the doubly degenerate $E_{2g}$ branch in the monolayer
evolves into two (three) doubly degenerate branches for bilayer
(trilayer) graphene, as shown in Fig.~\ref{fig:layer-ph}(b). These
small splittings are due to the weak interlayer coupling: about 5
cm$^{-1}$ for bilayer and no more than 5 cm$^{-1}$ for trilayer (see
Table~\ref{tab:freq}). Moving away from $\Gamma$, each of these
degenerate branches breaks into two different modes. Recent
experiments show that the Raman G-peak intensity enhances almost
linearly with respect to the layer number (up to four
layers).\cite{Graf2006,Gupta2006} This phenomenon could be ascribed
to the increased number of optical phonon modes at $\Gamma$ within a
small energy window for multilayer graphene.


The stackings of graphene layers have various point group symmetry
for the $\Gamma$ phonons. The monolayer graphene possesses the
$D_{6h}$ symmetry (Sch\"{o}nflies notation). It reduces to $D_{3d}$
for the AB bilayer and ABC trilayer, and $D_{3h}$ for the ABA
trilayer. Correspondingly, their high optical zone-center modes are
of different mode symmetry: $E_{2g}$ mode in graphene evolves into
$E_g$ and $E_u$ for the AB bilayer, 2$E'$+$E''$ for the ABA
trilayer, and 2$E_g$+$E_u$ for the ABC trilayer. The $E_g$ and $E''$
modes are Raman active, $E_u$ is IR active, while the $E'$ modes are
both Raman and IR active. Therefore, a complete picture of the
zone-center modes can be obtained from a combination of Raman and IR
measurements. These mode splittings provide significant information
about the layer number and the stacking geometry.

\begin{figure}[tbp]
\centering
   \includegraphics[width=6.5cm]{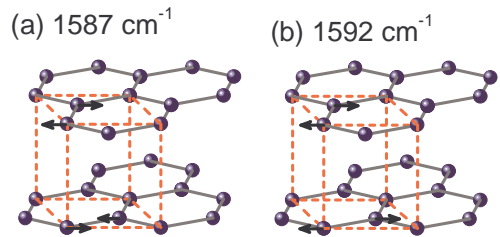}
 \caption{(Color online) Atomic displacements of the two split optical branches (a) 1587, and
 (b) 1592 cm$^{-1}$ at $\Gamma$ for the AB bilayer. Only one mode for each degenerate pair
 is shown. }\label{fig:AB-gamma}
\end{figure}

\begin{figure}[tbp]
\centering
   \includegraphics[width=8.5cm]{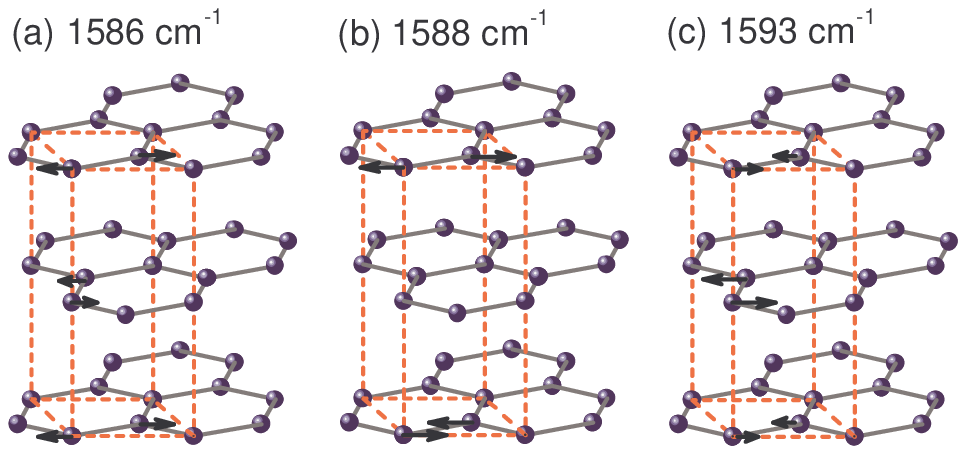}
 \caption{(Color online) Atomic displacements of the three split optical branches (a) 1586,
 (b) 1588, and (c) 1593 cm$^{-1}$ at $\Gamma$ for the ABA trilayer.
 The length of the arrow represents the amplitude of the eigenvector. Only one mode
 for each degenerate pair is shown. }\label{fig:ABA-gamma}
\end{figure}

\begin{figure}[tbp]
\centering
   \includegraphics[width=8.5cm]{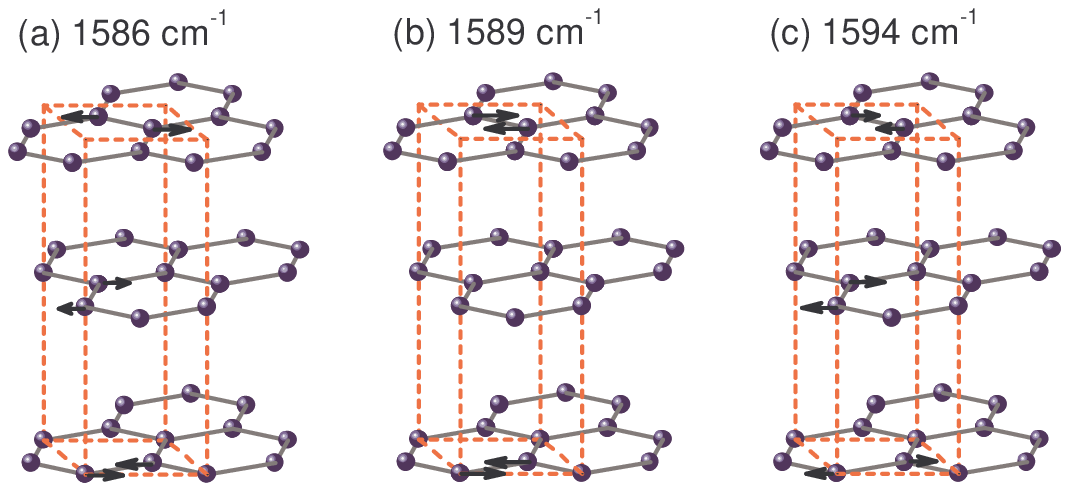}
 \caption{(Color online) Atomic displacements of the three split optical branches (a) 1586, (b) 1589,
 and (c) 1594 cm$^{-1}$ at $\Gamma$ for the ABC trilayer. The length of the arrow represents
 the amplitude of the eigenvector. Only one mode
 for each degenerate pair is shown. }\label{fig:ABC-gamma}
\end{figure}

In Figs.~\ref{fig:AB-gamma}-\ref{fig:ABC-gamma}, we show the
schematic atomic displacements of these optical eigenmodes at
$\Gamma$ for the AB bilayer, ABA trilayer, and ABC trilayer,
respectively. These high-frequency phonons are derived from the
superpositions of intralayer optical modes in each graphene plane.
For the modes in the bilayer as shown in Fig.~\ref{fig:AB-gamma},
the two atoms on top of each other in two adjacent layers vibrate
either in the opposite direction ($E_g$ mode, 1587 cm$^{-1}$) or in
the same direction ($E_u$ mode, 1592 cm$^{-1}$). Similar atomic
displacements can also be seen in ABA and ABC trilayers, as shown in
Figs.~\ref{fig:ABA-gamma} and \ref{fig:ABC-gamma}. In other words,
the original intralayer modes couple to each other via interlayer
interactions, giving rise to a small splitting in the final
frequencies. The upper and lower modes in the bilayer correspond to
the `in-phase' and `out-of-phase' superpositions of the two
intralayer modes, respectively, similar to the $E_{1u}$ and $E_{2g}$
modes in bulk graphite.

This splitting of the phonon frequencies at $\Gamma$ can be
illustrated using a simple model. Using the original intralayer
optical modes as the basis and assuming the interaction strength
between adjacent layers is $\epsilon$, the reduced Hamiltonian for
the bilayer and trilayer can be expressed as:
\begin{equation}
H_2=E_0 I + \left(
            \begin{array}{cc}
              0 & \epsilon \\
              \epsilon & 0 \\
            \end{array}
          \right)
 =\left(
            \begin{array}{cc}
              E_0 & \epsilon \\
              \epsilon & E_0 \\
            \end{array}
          \right)\label{eq:H2}
\end{equation}
and
\begin{equation}
H_3=\left(
            \begin{array}{ccc}
              E_0 & \epsilon & 0\\
              \epsilon & E_0+\delta & \epsilon\\
              0 & \epsilon & E_0 \\
            \end{array}
          \right)\label{eq:H3},
\end{equation}
respectively. Here, $E_0$ is the energy of the intralayer mode, and
only the first nearest-neighbor layer-layer interaction is
considered. For the trilayer, a small variant of $\delta$ is
introduced in Eq.~(\ref{eq:H3}) to account for the change of the
on-site energy in the middle layer due to the new geometry. (This is
similar to the on-site energy variation due to environmental changes
in electronic tight-binding models.\cite{Chadi1989,Mercer1994})
Solving the secular equation $\mathrm{det}(H-\lambda I)$=0, one
obtains the eigenvalues and eigenvectors.

For the bilayer, the eigenvalues are $\lambda_{1,2}=E_0\pm\epsilon$.
From Fig.~\ref{fig:AB-gamma}, the $E_0$ and $\epsilon$ can be
determined: $E_0$=1589.5 and $|\epsilon|$=2.5 cm$^{-1}$. $E_0$ shows
a small shift compared with the value for a single-layer graphene
(1586 cm$^{-1}$) as a result of the environmental  change mentioned
above. With $\epsilon>0$, the corresponding eigenvectors are
$\phi_{1,2}=(1, \pm1)^T$. This is consistent with the displacements
we obtained in Fig.~\ref{fig:AB-gamma}. The lower frequency
corresponds to the out-of-phase superposition of the two intralayer
modes (with respect to the motion of the two atoms on top of each
other in two adjacent layers), while the higher one corresponds to
the in-phase superposition. In both cases, the two intralayer modes
have equal amplitudes.

For the trilayer, the eigenvalues are $\lambda_1$=$E_0$,
$\lambda_{2,3}$=$E_0+(\delta\pm\sqrt{\delta^2+8\epsilon^2})/2$, with
corresponding eigenvectors $\phi_1$=$(1, 0, -1)^T$,
$\phi_{2,3}$=$(\epsilon,(\delta\pm\sqrt{\delta^2+8\epsilon^2})/2,\epsilon)^T$.
For the latter two eigenvectors, the ratio of the mode amplitudes in
each layer is
$a_1:a_2:a_3$=$\epsilon:(\delta\pm\sqrt{\delta^2+8\epsilon^2})/2:\epsilon$.
Using the frequencies for the ABA trilayer as shown in
Fig.~\ref{fig:ABA-gamma}, we obtain $\delta\approx3$~cm$^{-1}$ and
$\epsilon\approx2.2$~cm$^{-1}$, and the mode amplitude ratios of
2.2:5.0:2.2 and 2.2:(-2.0):2.2 for these two modes, respectively.
This result agrees with the displacements from the direct
first-principles calculations shown in Fig.~\ref{fig:ABA-gamma}.
Similar results can be obtained for the ABC trilayer. Based on these
numerical results, one can easily estimate the optical phonon
frequencies for more graphene layers using an interlayer interaction
of 2-3 cm$^{-1}$ and the values of $E_0$ and $\delta$ obtained
above.

The highest optical phonon branch at K becomes doubly degenerate at
1318 cm$^{-1}$ in the bilayer system (see Table~\ref{tab:freq}),
nearly 12 cm$^{-1}$ higher than that in graphene. The degeneracy is
imposed by the symmetry of the bilayer. The two degenerate modes
correspond to two intralayer modes within individual layers with
little coupling between them. In contrast, the optical phonons in
the ABA split into three modes, with two of them being almost
degenerate, while the degeneracy is imposed by symmetry in the ABC
trilayer. For the ABA and ABC trilayers, the highest phonon
(singlet) frequencies are 1324 and 1325 cm$^{-1}$ , respectively.
This result is consistent with the Raman observation that the
second-order D mode at about 2700 cm$^{-1}$ increases with an
increasing layer number. \cite{Ferrari2007} The second-order D mode
in the Raman spectrum of graphene and graphene layers, which is
double of the highest optical phonon frequency at K, can be well
illustrated using a double-resonant model.\cite{Ferrari2007}

\begin{figure}[tbp]
\centering
   \includegraphics[width=8.5cm]{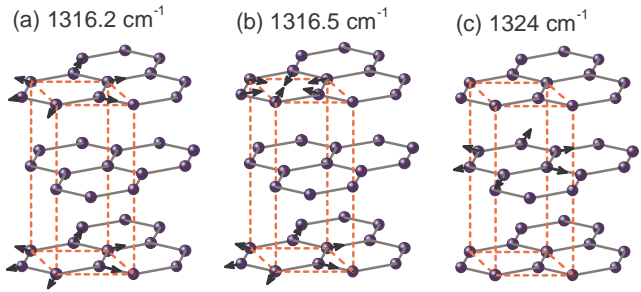}
 \caption{(Color online) Atomic displacements of the three optical phonon modes (a) 1316.2, (b) 1316.5, and (c) 1324 cm$^{-1}$ at K for the ABA trilayer.
 The length of the arrow represents the amplitude of the eigenvector. Only one mode
 for each degenerate pair is shown.}\label{fig:ABA-K}
\end{figure}

Figure~\ref{fig:ABA-K} shows the schematic atomic displacements of
the three eigenmodes at K for the ABA trilayer. Each eigenmode
comprises a superposition of the intralayer $A_1^{'}$ modes from
each layer. As shown in Fig.~\ref{fig:ABA-K}, the two almost
degenerate low-frequency modes (about 1316 cm$^{-1}$) correspond to
the combinations of the modes from the top and bottom A layers,
while the high branch is an intralayer mode from the middle B layer
almost exclusively.

The splitting in frequencies at K can be analyzed in a similar way
as before. In contrast to the $\Gamma$ phonons, the coupling between
adjacent intralayer modes is zero due to the mode symmetry, and
there might only be a small interaction between the top and the
bottom layers. In this case, the Hamiltonian can be expressed as:
\begin{equation}
H=\left(
            \begin{array}{ccc}
              E_0' & 0 & \eta\\
              0 & E_0'+\delta' &0 \\
              \eta & 0 & E_0' \\
            \end{array}
          \right),
\end{equation}
with $\eta$ the small interaction between second nearest-neighbor
layers. The eigenvalues are $\lambda=E_0'\pm\eta$, and
$E_0'+\delta'$. According to the frequencies as shown in
Fig.~\ref{fig:ABA-K}, we obtain $\delta'$=7.2 cm$^{-1}$ and
$|\eta|$=0.2 cm$^{-1}$. The amplitudes are $a_1:a_2:a_3$=$\pm1:0:1$
for $\lambda=E_0'\pm\eta$, and $0:1:0$ for $\lambda=E_0'+\delta'$.
This is consistent with the atomic displacements illustrated in
Fig.~\ref{fig:ABA-K} with $\eta<$0. Note that the small interactions
between second nearest-neighbor layers have induced a small
splitting of the two low-frequency modes.

In the case of ABC stacking, the interlayer coupling matrix elements
are identically zero. Therefore, the three eigenmodes at K are
localized on each of the three layers, respectively. Due to the
higher on-site energy in the middle layer, one state will be higher
than the other two modes, with the lower two modes doubly degenerate
due to the symmetry.

\section{Summary}

In summary, we have studied the phonon dispersions and vibrational
properties for monolayer, bilayer, and trilayer graphene using the
density-functional perturbation theory. Due to the weak coupling
between layers, the highest optical phonon branch at $\Gamma$ in
graphene splits into two (three) doubly degenerate branches with
small yet unnegligible splittings for bilayer (trilayer) graphene.
The splitting is about 5 cm$^{-1}$ for the bilayer. In trilayer ABA
and ABC graphene, these splittings are about 2 cm$^{-1}$ and 5
cm$^{-1}$, respectively, which are not equally spaced. These modes
are of various mode symmetry and exhibit different sensitivity to
either Raman or IR measurements and therefore a combination of Raman
and IR measurements of the zone-center optical modes should give a
clear identification of the layer number as well as the stacking
geometry.

A simple interaction model is applied to illustrate the frequency
splitting and the characteristics of the eigenmodes at $\Gamma$. The
interlayer coupling strength is identified as about 2 cm$^{-1}$. In
the trilayer system, a shift of about 3 cm$^{-1}$ in the on-site
energy in the middle layer is determined.

The frequency of the highest optical phonon mode at K in bilayer
(trilayer) graphene is about 12 (18) cm$^{-1}$ higher than that in
monolayer graphene. For trilayer graphene, the K-A$_1'$ mode splits
into three branches in the ABA trilayer, with the two lower modes
nearly doubly degenerate. It is found that the on-site energy
variations for the middle layer in ABA and ABC are about 7-8
cm$^{-1}$, higher than that of $\Gamma$ phonons. Due to the
symmetry, the interlayer coupling between adjacent layers for these
intralayer modes is zero.

\begin{acknowledgments}

We acknowledge helpful discussions with M. Wierzbowska, S. Piscanec,
and A. C. Ferrari. We thank M. Mohr for providing the experimental
data of bulk graphite. This work is supported by the Department of
Energy (Grant No. DE-FG02-97ER45632) and by the National Science
Foundation (Grants No. DMR-02-05328). The computation used resources
of the National Energy Research Scientific Computing Center (NERSC),
which is supported by the U.S. Department of Energy (Grant No.
DE-AC03-76SF00098), and San Diego Supercomputer Center (SDSC) at
UCSD.

\end{acknowledgments}

\end{document}